\documentclass[letterpaper,twocolumn,10pt]{article}
\usepackage{usenix2020}

\usepackage{tikz}
\usepackage{amsmath}

\usepackage{times}
\usepackage{soul}
\usepackage{url}
\usepackage[utf8]{inputenc}
\usepackage[small]{caption}
\usepackage{graphicx}
\usepackage{booktabs}
\usepackage{algorithm}
\usepackage[noend]{algpseudocode}
\usepackage{amssymb,amsfonts}
\usepackage{amsthm}
\usepackage{stmaryrd}
\usepackage{multirow}
\usepackage{mathtools}
\usepackage{pifont}

\usepackage{mdframed}
\newmdtheoremenv{prob}{Problem}
\newmdtheoremenv{defi}{Definition}

\newcommand{\sample}{\ {\leftarrow}{\mbox{\tiny \$}} \ }

\begin{document}
%-------------------------------------------------------------------------------

%don't want date printed
%\date{}

% make title bold and 14 pt font (Latex default is non-bold, 16 pt)
\title{Popcorn: Paillier Meets Compression For Efficient Oblivious Neural Network Inference}

%for single author (just remove % characters)
% \author{
% {\rm Jun Wang}\\
% Post, Luxembourg\\
% % Institute for Infocomm Research, A*STAR\\
% junwang.lu@gmail.com 
% \and
% {\rm Chao Jin}\\
% Institute for Infocomm Research, A*STAR\\
% jin\_chao@i2r.a-star.edu.sg
% \and
% {\rm Souhail Meftah}\\
% Institute for Infocomm Research, A*STAR\\
% souhail\_meftah\_from.tp@i2r.a-star.edu.sg
% \and
% {\rm Khin Mi Mi Aung}\\
% Institute for Infocomm Research, A*STAR\\
% mi\_mi\_aung@i2r.a-star.edu.sg
% } % end author
\author{
{\rm Jun Wang$^1$, Chao Jin$^2$, Souhail Meftah$^2$, Khin Mi Mi Aung$^2$}\\
$^1$Post, Luxembourg\\
$^2$Institute for Infocomm Research, A*STAR\\
{\rm junwang.lu@gmail.com, jin\_chao@i2r.a-star.edu.sg}
}
\maketitle

%-------------------------------------------------------------------------------
\begin{abstract}
%-------------------------------------------------------------------------------
Oblivious inference enables the cloud to provide neural network inference-as-a-service (NN-IaaS), whilst neither disclosing the client data nor revealing the server's model. However, the privacy guarantee under oblivious inference usually comes with a heavy cost of efficiency and accuracy.

We propose Popcorn, a concise oblivious inference framework entirely built on the Paillier homomorphic encryption scheme. We design a suite of novel protocols to compute non-linear activation and max-pooling layers. We leverage neural network compression techniques (i.e., neural weights pruning and quantization) to accelerate the inference computation. To implement the Popcorn framework, we only need to replace algebraic operations of existing networks with their corresponding Paillier homomorphic operations, which is extremely friendly for engineering development.

We first conduct the performance evaluation and comparison based on the MNIST and CIFAR-10 classification tasks. Compared with existing solutions, Popcorn brings a significant communication overhead deduction, with a moderate runtime increase. Then, we benchmark the performance of oblivious inference on ImageNet. To our best knowledge, this is the first report based on a commercial-level dataset, taking a step towards the deployment to production.
\end{abstract}

%-------------------------------------------------------------------------------
\section{Introduction}
%-------------------------------------------------------------------------------

%1.	Why inference as a service, 
Deep convolutional neural networks have achieved tremendous success in various domains such as facial recognition~\cite{schroff2015facenet}, medical diagnosis~\cite{esteva2017dermatologist}, and image classification~\cite{he2016deep}, whilst demonstrating beyond-experts performance. The massive (labeled) training data and extensive computational resources are the fuel for breakthroughs in accuracy. However, it also becomes a notorious challenge for individuals and non-specialised institutions to train and deploy state-of-the-art models. Thanks to the advances in cloud computing, companies with sufficient computing power and expertise (e.g., Google and Amazon) can provide machine learning services to the public. NN-IaaS is an important business paradigm, in which a server provides machine learning prediction APIs, built upon its pre-trained machine learning model, to the clients. The latter uploads her data to the server and receives predictions by calling the APIs.

%2. why oblivious inference
Usually, business and privacy protection requirements may prevent the server from providing any information other than prediction results. Similarly, the clients are protective of their private data and cannot disclose it to the server. This dilemma significantly limits the use of cloud-based services, leading to an urgent need for privacy-preserving NN-IaaS.

%2.	State-of-the-art, information leakage.
Homomorphic encryption (HE)~\cite{gentry2009fully}, Garbled Circuits (GC)~\cite{bellare2012foundations} and secret sharing (SS)~\cite{beimel2011secret} are the workhorse driving many exciting recent advances in oblivious neural network inference~\cite{gilad2016cryptonets,liu2017oblivious,riazi2019xonn,juvekar2018gazelle}. CryptoNets~\cite{gilad2016cryptonets} and its variant~\cite{chou2018faster} adopted HE to support privacy-preserving neural network predictions. Since HE operations are constrained to addition-and-multiplications, the non-linear activation function (e.g., relu$(x,0)$) is substituted with a low-degree polynomial; and the max-pooling function is replaced by the mean-pooing function. This approach requires modifying the original model architecture, significantly impacting accuracy. XONN~\cite{riazi2019xonn} exploited the fact that the XNOR operation is free in the GC protocol~\cite{kolesnikov2008improved} to efficiently evaluate binarized neural networks. This approach, however, requires both the weights and activation values to be binarized, which harms accuracy performance. Moreover, completely compiling a neural network into circuits increases communication overhead.  MiniONN~\cite{liu2017oblivious} and Gazelle~\cite{juvekar2018gazelle} combined different primitives to keep neural networks unchanged. In their methods, the GC is often adopted to compute non-linear layers. It is worth highlighting that existing methods always leak some information beyond the predictions. For example, in CryptoNets, the client can make inferences
about the model, as it would have to generate parameters for the encryption according to the model architecture. In XONN and MiniONN, the client can learn the exact network architecture. Gazelle only reveals the number of neurons of each neural layer. Though there is always some form of information leakage, it is believed that less leakage is more preferable~\cite{juvekar2018gazelle,riazi2019xonn}. Engineering complexity is also an important problem to be addressed. For instance, Gazelle relies on sophisticated packing schemes. Its implementation depends on specific network parameters. Sometimes, it also requires performing trade-offs between efficiency and privacy. XONN needs to compile a whole network into circuits, the effort is non-trivial; the computational cost is also completely transferred from the server to the client.

%Z
%We propose a lightweight protocol,
In this paper, we introduce Popcorn, a concise oblivious inference framework that is entirely built upon the Paillier homomorphic encryption scheme~\cite{paillier1999public}. Popcorn is a non-invasive solution which does not require modifications to network architectures (e.g., approximating non-linear activation functions with polynomials). The security model of Popcorn is consistent with Gazelle~\cite{juvekar2018gazelle}, i.e., this protocol hides the network weights and architecture except for the number of neurons of each layer. The main contribution of this paper consists of three aspects,
\begin{itemize}
    \item We introduce a suite of Popcorn protocols for efficiently computing non-linear neural network layers (e.g., $relu$ activation layer and max-pooling layer).
    \item Under the Popcorn framework, we leverage network compression (e.g., weight pruning and quantization) to accelerate the computation of linear layers (e.g., convolutional layer and fully-connected layer).
    \item We benchmark the oblivious inference performance on the ImageNet dataset, using state-of-the-art models. To our best knowledge, this is the first report on privacy-preserving ImageNet-scale classification tasks.
\end{itemize}

Compared with existing solutions,  an important contribution of Popcorn is that its engineering is extremely simple.  The framework completely relies on the Paillier HE scheme. In the implementation, we only need to substitute the algebraic operations in plaintext inference to the corresponding HE operations, making (machine learning) engineers agnostic to the obscure cryptography knowledge. Although the framework is built on the Paillier scheme in this paper, we can directly adopt other HE schemes. For example, in the case where a client has a number of images (e.g., $>$1000) to classify, we can also adopt a lattice-based additive HE scheme (e.g., ~\cite{juvekar2018gazelle}) to pack multiple images into one ciphertext, amortizing the computational and communicational overhead.

\section{Preliminaries}
\subsection{Convolutional Neural Network}
A typical convolutional neural network (CNN) consists of a sequence of linear and non-linear layers and computes classifications in the last layer. In this section, we describe the functionality of neural network layers.

\subsubsection{Linear Layers}

Linear operations in networks are often carried in fully-connected layers ($fc$) and  convolutional layers ($conv$). 

The $conv$ layer is composed of a 3D tensor input (in the form of $\mathbb{R}^{(w_i, h_i, c_i)}$), a set of 3D tensor filters (in the form of $\mathbb{R}^{(f_w, f_h, f_c)}$, s.t. $f_w<w_i, f_h<h_i, f_c=c_i$) to extract local features from the input and a 3D tensor output (in the form of $\mathbb{R}^{(w_o, h_o, c_o)}$, $c_o$ is the number of channels of the output, which is also the number of filters). Each channel of the output is obtained by a filter that convolves the input along the direction of $w_i$ and $h_i$ (i.e., the width and height of the input). Specifically, each element of a channel is calculated through filter point-wisely multiplicating its perceptive field of the input and accumulating. The $fc$ layer is a matrix-vector multiplication as follows, 
\begin{equation}
\label{eq:mma}
 \textbf{y}=\textbf{W}\cdot\textbf{x}+\textbf{b}   
\end{equation}
where $\textbf{x} \in \mathbb{R}^{m\times1}$ is the input vector, $\textbf{y}\in \mathbb{R}^{n\times1}$ is the output,$\textbf{W}\in \mathbb{R}^{n\times m}$ is the weight matrix and $\textbf{b}\in \mathbb{R}^{n\times1}$is the bias vector. In fact, a $conv$ layer can be also written in a form of matrix-vector multiplication, as defined in Eq.~(\ref{eq:mma}).

In addition to $conv$ and $fc$ layers, the batch-normalization ($bn$)~\cite{ioffe2015batch} layer can be seen as a linear layer. It is often adopted to normalize the output of a linear layer by re-centering and re-scaling as follows,
 \begin{equation}
 \begin{split}
    y & =\gamma \cdot \frac{x-\mu_B}{\sqrt{\delta^2_B+\epsilon}}+\beta \\
    & =   \frac{\gamma}{\sqrt{\delta^2_B+\epsilon}}\cdot x - (\frac{\gamma\cdot \mu_B}{\sqrt{\delta^2_B+\epsilon}}-\beta)
 \end{split}
 \end{equation}
where $\mu_B$ and $\delta^2_B$ are the per-dimension mean and variance of a mini-batch, respectively. $\epsilon$ is a is an arbitrarily small constant added in the denominator for numerical stability. The $\gamma$ and $\beta$ are the re-scaling and re-centering parameters subsequently learned in the optimization process. In the inference phase, all the parameters above are constant. Therefore, the $bn$ computation is a linear computation in the inference and can be easily absorbed in its previous layer and next layer~\cite{ibarrondo2018fhe}. %In this paper, we skip the discussion of how to optimize the $bn$ computations. 

\subsubsection{Non-linear Layers}

The activation layer and pooling layer are two common non-linear layers. The activation layer performs non-linear transformation for its inputs element-wise. It does not change the input dimension. The $relu$ (i.e., $\max(x,0)$) activation function is widely adopted as the non-linear transformation function. Different from the activation layer, the pooling layer is for feature dimension reduction. The max-pooling ($mp$) is a popular pooling method, it outputs the maximal value of each local perception area  of its inputs. Suppose the local perception area has $m$ inputs, the max-pooling outputs the result of $\max(\{x_0, x_1,\cdots,x_m\})$. There is also another pooling operation, i.e., mean-pooling, it outputs the average value of each local perception area. The mean pooling layer can be treated as a special simplified convolution layer with a filter in which the weights share the same value. In practice, the mean pooling can be further simplified and replaced with sum-pooling, which outputs the sum of the perception area.

%The classifier is the last layer of a neural network, it outputs the final prediction results of a data sample. The softmax ($y=$)

\subsection{Neural Network Compression}
% what is network compression
To accelerate neural network predictions and minimize network size, the compression technique is developed to discover and remove unimportant weights (i.e., network pruning), and to present weights with fewer bits (i.e., weights quantization), without noticeably decreasing accuracy performance~\cite{yang2017designing,han2015deep}. 

\emph{Network Pruning.} Weight pruning and filter pruning (a.k.a channel pruning) are two main network pruning methods~\cite{yang2017designing}. The former investigates the removal of each individual weight (fine-grained pruning). Normally, the weights that have a small magnitude or contribute less to the network loss function are removed first~\cite{han2015deep,lee2018snip}. The latter investigates removing entire filters (coarse-grained pruning). Usually, filters that frequently generate zero outputs after the $relu$ layer are removed first ~\cite{yang2017designing}. Generally, the weight pruning approach can remove more weights than the filter pruning approach does, but the filter pruning methods can remove more neurons (we call each element of any output layer a neuron). Though both the pruning approaches can benefit the Popcorn framework, we focus on the weight pruning methods in this paper. 

\emph{Weight Quantization.} The quantization methods aim to reduce the number of bits to represent neural weights, which can be roughly categorized into two approaches: floating-point preserving ~\cite{han2015deep} and integer-based~\cite{krishnamoorthi2018quantizing}. In the former, the weights are quantized into a small number of bins (e.g., 128), all the weights in the same bin share the same value. Thus, for each weight, we need to store only a small index into a table of shared weights. In the latter, the weights are first quantized into integer representations and then recovered into a more expressive form (e.g., floating-point or more levels) through a de-quantization process, during the inference phase~\cite{krishnamoorthi2018quantizing}. Binarized quantization is an extreme case, where the weights or activation values are represented in 1-bit~\cite{rastegari2016xnor}, with a cost of accuracy loss. In this work, we leverage the floating-point preserved approach and binarized neural networks to accelerate oblivious neural network inferences.

% classification of compression methods

%Under this scheme, given the public key ($pk$) and the encryption of two messages $m_1$ and $m_2$, one can compute the encryption of $m_1+m_2$. Moreover, the multiplication between a ciphertext and a plaintext is also enabled. Specifically,

\subsection{Paillier Homomorphic Encryption}
\label{sec:he}
Homomorphic encryption (HE) is a form of encryption that allows computations to be carried over ciphertexts without decryption. The result, after decryption, is the same as if the operations had been performed on the plaintexts. The Paillier Cryptosystem~\cite{paillier1999public} is a well-developed additive HE scheme. We describe the Paillier HE scheme in the form of $\{ $ \emph{KeyGen}, \emph{HEnc}, \emph{HAdd}, \emph{HMul}, \emph{HDec}  $\}$.  
%We denote the encryption of $m$ as $\llbracket m \rrbracket$.

\emph{ $\bullet$  KeyGen (Generate keys: $(pk, sk)$ ).} 
\begin{enumerate}
    \item Choose two large prime numbers $p$ and $q$ randomly and independently, such that $\gcd(pq,(p-1)(q-1))=1$. %The $\gcd$ stands for greatest common divisor.
    \item Compute $n=pq$ and $\lambda=lcm(p-1,q-1)$, where $lcm$ means least common multiple.
    \item Obtain public key $pk = n, g$; private key $sk = p, q, \lambda$. $g\in \mathbb{Z}^{*}_{n^2}$ is a multiple of $n$.
\end{enumerate}

\emph{$\bullet$  HEnc (Encryption: $\llbracket m \rrbracket$ := HEnc$(m, pk)$ ).} 
\begin{enumerate}
    \item Let $m$ be a message to be encrypted, where $0\leq m < n$.
    \item Select a random $r\in \mathbb {Z} _{n}^{*}$, s.t. $\gcd(r,n)=1$.
    \item Compute the ciphertext of $m$: $\llbracket m \rrbracket \leftarrow g^mr^n \mod n^2$.
\end{enumerate}

\emph{$\bullet$  HDec (Decryption: $m :=$ HDec$(\llbracket m \rrbracket, pk)$).} 
\begin{enumerate}
    \item Let $\llbracket m \rrbracket$ be a cipher to be decrypted, where $ \llbracket m \rrbracket < n^2$.
    \item Compute the plaintext of  $\llbracket m \rrbracket$: $m \leftarrow \frac{L(\llbracket m \rrbracket^{\lambda} \mod n^2)}{L(g^{\lambda} \mod n^2)} \mod n$, where $L(x)=\frac{x-1}{n}$. 
\end{enumerate}

\emph{$\bullet$  HAdd (The HE addition)}
\begin{enumerate}
    \item $m_1+m_2$ = \emph{HDec}$(\llbracket m_1 \rrbracket \oplus \llbracket m_2 \rrbracket, sk )$. $\oplus$ is the HE addition operator.
\end{enumerate}

\emph{$\bullet$  HAdd (The HE multiplication)  }
\begin{enumerate}
    \item $m_1\times m_2$ = \emph{HDec}$(\llbracket m_1 \rrbracket \otimes m_2, sk )$. $\otimes$ is the HE multiplication operator.
\end{enumerate}

\section{Popcorn: Fast Non-linear Computation}
% some introduction
In this section, we present our fast methods for non-linear layer computations. The computation protocols are built upon the Paillier HE scheme~\cite{paillier1999public}. %We emphasize that the protocols can be also implemented through other additive homomorphic encryption schemes.

% a protocol to compute relu
\subsection{Fast Secure $relu$ Computation Protocol}  

\begin{prob}
\label{prob:1}
Given a vector $\llbracket \textbf{x} \rrbracket^{m \times 1}$ which is element-wisely encrypted under the Paillier HE scheme (where  $x_i\in \mathbb{Z}_n^*$), the server computes $relu(\llbracket \textbf{x} \rrbracket^{m \times 1})$, without leaking the information of any element $x_{i}$. In this setting, the client holds the secret key $sk$, the server holds the public key $pk$ but cannot access to the $sk$.
\end{prob}

We describe the secure computation of $relu$ activation layer in Problem~\ref{prob:1}, presenting the inputs in form of vector. The $relu$ layer performs non-linear transformation on each element of its inputs, i.e., $\{ max(x_i,0) \}_{i=0}^{n-1}$. To solve the Problem~\ref{prob:1}, we start from a simple multiplicative-obfuscation protocol for each $x_i$ (named Version one), which is outlined in Fig.~\ref{tab:naive}. For denotation succinctness, we omit the subscript of $x_{i}$ and we assume the implementation of Paillier HE supports encoding of of negative integers.

\begin{figure}[h!]
\centering
\centering\framebox{
\begin{tabular}{llll}
%\hline
$\downarrow$ & \textbf{Server}  ($\llbracket x \rrbracket,pk$)                                                             &                                                                           &                                                                               \\ \hline
1&$\tau \sample \mathbb{Z}^+_n\ s.t.\ \gcd(\tau, n)=1 $&                                                                           &                                             \\
2&$\llbracket y  \rrbracket_s := \llbracket x \rrbracket \odot \tau$ &                                                                           &                                             \\
3&                                                                &  $\xrightarrow{\llbracket y  \rrbracket_s}$  &                                                                                       \\ \hline
$\downarrow$&       \textbf{ Client} ($\llbracket y  \rrbracket_s, pk, sk$)                                                           &  &                                                                                       \\ \hline
4&    $\quad y_s:=$\emph{HDec}$(\llbracket y  \rrbracket_s,sk)$                                                          &                                                                           & \\
5&  $\quad  y_c:=y_s$ if $y_s>0$ else $0$    &                                                                           &\\
6& $\quad  \llbracket y  \rrbracket_c :=$ \emph{HEnc}$(y_c, pk)$    &                                                                           & \\                                                                                                                                  
 7&                                                            &    $\xrightarrow{\llbracket y \rrbracket_c}$                         &                                                                                       \\ \hline
$\downarrow$& \textbf{Server} ($\llbracket y \rrbracket_c, pk$)                                                                &                             &                                                                                       \\ \hline
8&  $\llbracket y  \rrbracket_s := \llbracket y \rrbracket_c \odot \tau^{-1}$  &                                                                           &                                                                                       \\
\end{tabular}
}
\caption{Secure $relu$ computation (Version one). $\tau^{-1}$ is the multiplicative inverse of $\tau$. $n$ is a large modulo.}
\label{tab:naive}
\end{figure}

In the Version one, the server first samples a random positive integer $ \tau\in \mathbb{Z}_n^+$, then it blinds the $\llbracket x \rrbracket$ with $\tau$ (i.e., $\llbracket y \rrbracket_s := \llbracket x \rrbracket \odot \tau$), and sends it to the client. After receiving the $\llbracket y \rrbracket_s$, the client decrypts $\llbracket y \rrbracket_s$, and returns the server with $\llbracket max(y_s,0) \rrbracket$. The server removes the $\tau$ by multiplying $\tau^{-1}$, where $\tau^{-1}$ the inverse of $\tau$. We emphasize that $\tau$ is independently and randomly sampled for each element $x$. 

If $y_c=0$ (line 5), the correctness proof is exactly the same of the decryption of Paillier HE scheme. Therefore, we focus the correctness proof on $\llbracket y  \rrbracket_s := \llbracket y \rrbracket_c \odot \tau^{-1}$ (line 8), where $y_c \neq 0$, i.e., the decryption of $\llbracket y  \rrbracket_s$ is $x$. Clearly, if we can eliminate the random factor $\tau$ when performing decryption, it is a normal Paillier decryption process(see Section~\ref{sec:he}, \emph{HDec}) and the correctness is guaranteed.
\begin{proof}[Correctness Proof of Version one] 
To decrypt $\llbracket y  \rrbracket_s$, we need to compute $(\llbracket y  \rrbracket_s^\lambda \mod n^2)$ as follows
\begin{equation}
\nonumber
    \begin{split}
         \llbracket y  \rrbracket_s^{\lambda} &= (g^yr^n)^{\tau^{-1}{\lambda}} \mod n^2 \\ 
       &=g^{\lambda x\tau\tau^{-1}}r^{\lambda n\tau^{-1}} \mod n^2 \\
       &\equiv g^{\lambda x\tau\tau^{-1}} \mod n^2 \\
       &=(1+n)^{\lambda x\tau\tau^{-1}} \mod n^2\\
       &=1+nx\lambda \tau\tau^{-1} \mod n^2  \quad (\triangleright \tau\tau^{-1} \equiv 1 \mod n)\\
       &=1+nx\lambda \mod n^2
    \end{split}
\end{equation}
As above, the random factor $\tau$ has been eliminated by its multiplicative inverse $\tau^{-1}$, the rest is exactly the same with the decryption of the Paillier HE scheme. The decryption of $\llbracket y  \rrbracket_s$ is $x$, the correctness is established.
\end{proof}

As the proof above, with protocol Version one , the server correctly gets $max(\llbracket x \rrbracket,0)$ and learns nothing about $x$. However, since $\tau \in \mathbb{Z}_n^+$, the client learns two pieces of information about $x$, (1) the sign of $x$ and (2) whether $x=0$ or not (denoted as $x \stackrel{?}{=} 0$). In the rest of this subsection, we first introduce the method to hide the sign of $x$ then to disguise $x \stackrel{?}{=} 0$.

\subsubsection{Hide the Sign of $x$}
The root cause for revealing the sign information of $x$ is that the blind factor $\tau$ is restricted to be positive (i.e., $\tau\in \mathbb{Z}_n^+$). To hide the sign information of $x$, we revise version one of the protocol (see Fig.~\ref{tab:naive}) to allow the blind factor $\tau$ to be a nonzero integer (e.g., $\tau \in \mathbb{Z}_{n}^*, s.t.\ \gcd(\tau, n)=1$). We outline the new version, named Version two, in Fig.~\ref{tab:hidesign}.

\begin{figure}[h!]
\centering
\centering\framebox{
\begin{tabular}{llll}
%\hline
& \textbf{Server}  ($\llbracket x \rrbracket, pk $)                                                             &                                                                           &                                                                              \\ \hline
1& $\tau \sample \mathbb{Z}^*_{n}\ s.t.\ \gcd(\tau, n)=1$ &                                                                           &                                             \\
2&$\llbracket y  \rrbracket_s := \llbracket x \rrbracket \odot \tau$ &                                                                           &                                             \\
3&                                                                 & $\xrightarrow{\llbracket y  \rrbracket_s}$ &                                                                                       \\ \hline
$\downarrow$&   \textbf{Client}  ($\llbracket y  \rrbracket_s, pk, sk$)                                                                &  &                                                                                       \\ \hline
4&  $\quad y_s:=$\emph{HDec}$(\llbracket y  \rrbracket_s,sk)$                                     &                                                                           & \\
5&  $\quad y_c:=y_s$ if $y_s>0$ else $0$  &                                                                           & \\
6&  $\quad \llbracket y  \rrbracket_c := $\emph{HEnc}$(y_c,pk)$ &                                                                           &  \\                                                                                                                                  
 7&                                                                 & $\xrightarrow{\llbracket y \rrbracket_c}$                             &                                                                                       \\ \hline
$\downarrow$&      \textbf{Server} ($\llbracket y \rrbracket_c, pk$)                                                           &                             &                                                                                       \\ \hline
8& $\llbracket y  \rrbracket_s := \llbracket y \rrbracket_c \odot \tau^{-1}$  &                                                                           &                                                                                       \\
9& if $\tau<0$:  &                                                                           &                                                                                       \\
10& \quad $\llbracket y  \rrbracket_s := \llbracket x \rrbracket  - \llbracket y \rrbracket_s$  &                                                                           &                                                                                       \\
%\Comment{post-processing $\hat{\textbf{r}}_v$}   &                                                                           &                                                                                       %\\ \hline
\end{tabular}
}
\caption{Version two: hide the sign information of $x$. }
\label{tab:hidesign}
\end{figure}

Compared to the protocol Version one (Fig.~\ref{tab:naive}), the operations at the client side remain the same. The key difference is that when the server receives the response (i.e., $\llbracket y  \rrbracket_c$) from the client, the server  computes the $relu$ activation depending on the blind factor $\tau$. If $\tau>0$, the $relu$ calculation is the same with version one of the protocol; if $r<0$, it means the client always provides the server the opposite information (e.g., when $x>0$, the client returns $\llbracket 0 \rrbracket$.). Therefore, the server computes the $relu$ in an opposite way, $\llbracket y  \rrbracket_s := \llbracket x \rrbracket  - \llbracket y \rrbracket_c \odot \tau^{-1}$.  If $\tau >0$, the proof is the same as Version one. Here, we focus the proof of Version two on $\tau < 0$.

\begin{proof}[Correctness Proof of Version two] 
$\newline$
If $x>0$, the client returns $\llbracket y \rrbracket_c = \llbracket 0 \rrbracket$, the server computes $\llbracket y  \rrbracket_s := \llbracket x \rrbracket - \llbracket 0 \rrbracket \odot \tau^{-1}$, obtaining $\llbracket y  \rrbracket_s = \llbracket x \rrbracket$ as the expectation (i.e., $\max(x,0)$).
$\newline$
If $x\leq 0$, the client returns $\llbracket y \rrbracket_c = \llbracket x\cdot \tau \rrbracket$ (line 5, $y_s=x\cdot \tau$), the server computes $\llbracket y  \rrbracket_s := \llbracket x \rrbracket - \llbracket x\cdot \tau \rrbracket \odot \tau^{-1}$, getting $\llbracket y  \rrbracket_s = \llbracket 0 \rrbracket$ as the expectation. The correctness is established.
\end{proof}
 
\subsubsection{Hide $x_i \stackrel{?}{=} 0$}
Straightforwardly applying the protocol Version two (see~\ref{tab:hidesign}) to Problem~\ref{prob:1} leaks $x_i \stackrel{?}{=} 0$. We address this issue by randomly shuffling the input elements feeding to the $relu$ layer. After the shuffling, the client cannot trace where a value was originally placed (i.e., un-traceable), thus to hide  whether the true value of a specific slot is zero or not. By this, the client only learns the number of zero values, which can be hidden by adding dummy elements. Formally, we describe the un-traceability of $x_i\in \textbf{x}^{m\times 1} $  in the Definition~\ref{def:untrac}.

\begin{defi}
\label{def:untrac}
The location of $x_i \in \textbf{x}^{m\times 1}$ is un-traceable if an observer is unable to distinguish $y_i \in \textbf{y}^{m\times 1}$ from a random, where $y_i = x_i'\cdot \tau_i$ and  $x_i'$ is the value at slot $i$ of $\textbf{x}^{m\times 1}$, not $x_i$ itself (through randomly shuffling).
\end{defi}

To achieve the un-traceability, the server needs a pair of uniform-random shuffling functions ($\Pi, \Pi^{-1}$) as follows, 
\begin{equation}
 \begin{split}
    & \llbracket \textbf{x}' \rrbracket^{m\times 1} \leftarrow \Pi (\llbracket \textbf{x} \rrbracket^{m\times 1}, \gamma)\\
    &  \llbracket \textbf{x} \rrbracket^{m\times 1} \leftarrow \Pi^{-1} (\llbracket \textbf{x}' \rrbracket^{m\times 1}, \gamma)
\end{split}     
\end{equation}
where the $\gamma$ is a private random seed. 

Specifically, given an input vector $\llbracket \textbf{x} \rrbracket^{m\times 1}$, the server first adds $t$ dummy elements (of which a random portion are set to 0) to the input vector, and gets $\llbracket \textbf{x} \rrbracket^{(m+t)\times 1}$. This step is to hide the number of elements with value of 0. For denotation succinctness, we let $m=m+t$, i.e., we continue to denote the new input vector as $\llbracket \textbf{x} \rrbracket^{m\times 1}$.  The second, the server samples a private random seed $\gamma$ and applies the random shuffling function $\Pi$ to the input  vector $\llbracket \textbf{x} \rrbracket^{m\times 1}$, obtaining  $\llbracket \textbf{x}' \rrbracket^{n\times 1}$. The third, the server blinds each $\llbracket x_i' \rrbracket \in \llbracket \textbf{x}' \rrbracket$ with a independently and randomly sampled integer $\tau \in \mathbb{Z}_{n}^*\ s.t. \ \tau \neq 0$ ( $\llbracket y_i \rrbracket_s = \llbracket x_i' \rrbracket \cdot \tau_i$, see Fig.~\ref{tab:hidesign}). With the random shuffling and one-time random mask (i.e., $\tau_i$), it is clearly that $y_i$ is indistinguishable from a random value (i.e., $x_i\in \textbf{x}^{n\times 1} $ is un-traceable), thus to hide the sign information of $x_i$. We emphasize that, after applying the shuffling function $\Pi$ to the input array $\textbf{x}^{m\times 1}$, $x_i'$ is the value at slot $i$ ($y_i = x_i'\cdot \tau_i $), not $x_i$ itself.

\begin{figure}[h]
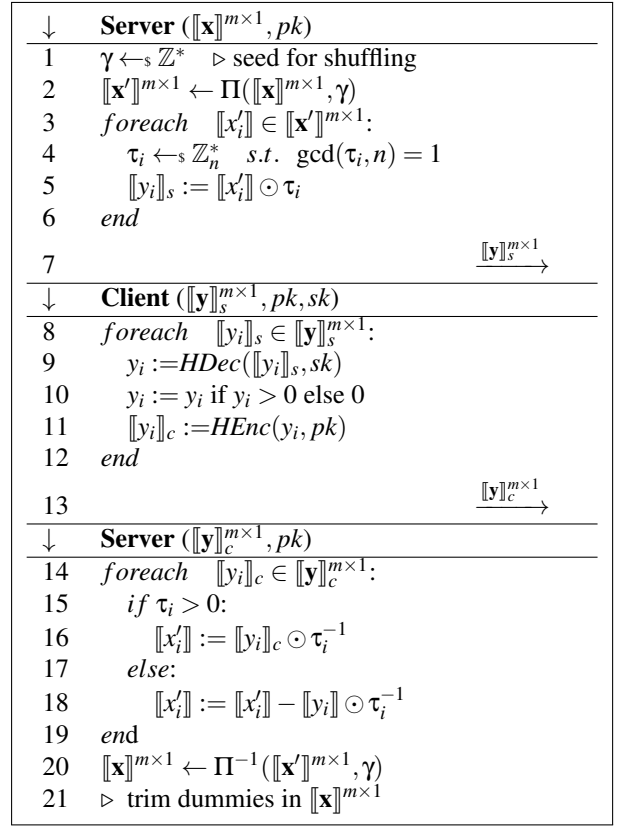

\centering
\centering\framebox{
\begin{tabular}{llll}
%\hline
$\downarrow$& \textbf{Server}  ($\llbracket \textbf{x} \rrbracket^{m \times 1}, pk $)                                                             &                                                                           &                                                                               \\ \hline
1& $\gamma \sample \mathbb{Z}^* \quad \triangleright $ seed for shuffling  &                                                                           &                                             \\
2&$\llbracket \textbf{x}' \rrbracket^{m \times 1} \leftarrow \Pi (\llbracket \textbf{x} \rrbracket^{m \times 1}, \gamma) $ &                                                                           &                                             \\
3&$foreach \quad \llbracket x_i' \rrbracket \in \llbracket \textbf{x}' \rrbracket^{m \times 1}$: &                                                                           &                                             \\
4&$\quad \tau_i \sample \mathbb{Z}_n^* \quad s.t. \ \ \gcd(\tau_i, n)=1$ &                                                                           &                                             \\
5&$\quad \llbracket y_i  \rrbracket_s := \llbracket x_i' \rrbracket \odot \tau_i$ &                                                                           &                                             \\
6& $end$ &                                                                           &                                             \\ 
7&            &  $\xrightarrow{\llbracket \textbf{y}  \rrbracket_s^{m \times 1}}$  &                                                                                       \\ \hline 
$\downarrow$ &    \textbf{Client} ($\llbracket \textbf{y}  \rrbracket_s^{m \times 1}, pk, sk$)         &  &                                                                                       \\ \hline 
8& $ foreach \quad \llbracket y_i \rrbracket_s \in \llbracket \textbf{y} \rrbracket_s^{m \times 1}$: &                                    &                                             \\
9& $\quad y_i:=$\emph{HDec}$(\llbracket y_i  \rrbracket_s,sk)$ &                                                                           &  \\
10& $\quad y_i:=y_i$ if $y_i>0$ else $0$ &                                                   &  \\
11& $\quad \llbracket y_i  \rrbracket_c := $\emph{HEnc}$(y_i,pk)$&                                   &  \\  
12& $end$ &                               & \\                                                                                                                                   
13&        &       $\xrightarrow{\llbracket \textbf{y} \rrbracket_c^{m \times 1}}$                      &                                                                                       \\ \hline
$\downarrow$&  \textbf{Server} ($\llbracket \textbf{y} \rrbracket_c^{m \times 1}, pk$)     &                            &                                                                                       \\ \hline
14&  $foreach \quad \llbracket y_i \rrbracket_c \in \llbracket \textbf{y} \rrbracket_c^{m \times 1}$:   &                                                                           &                                                                                       \\
15& \quad $if$ $\tau_i>0$:   &                                                                           &                                                                                       \\
16& \quad \quad  $\llbracket x_i'  \rrbracket := \llbracket y_i \rrbracket_c \odot \tau_i^{-1}$  &                                                                           &                                                                                       \\
17& \quad $else$:  &                                                                           &                                                                                       \\
18& \quad \quad $\llbracket x_i'  \rrbracket := \llbracket x_i' \rrbracket  - \llbracket y_i \rrbracket \odot \tau_i^{-1}$  &                                                                           &                                                                                       \\
19& $en$d  &                                                                           &                                                                                       \\
20& $\llbracket \textbf{x} \rrbracket^{m\times 1} \leftarrow \Pi^{-1} (\llbracket \textbf{x}' \rrbracket^{m \times 1}, \gamma)$   &                                                                        &                                                                                       \\
21& $\triangleright \ $ trim dummies in  $\llbracket \textbf{x} \rrbracket^{m\times 1} $  &                                                                        &                                                                                       \\
\end{tabular}
}
\caption{Secure $relu$ computation (Complete version). We assume the $\llbracket \textbf{x} \rrbracket^{m \times 1}$ already contains dummies. }
\label{tab:completedrelu}
\end{figure}

Now, we have introduced the completed version of our secure $relu$ computation protocol, outlined in Fig.~\ref{tab:completedrelu}.

\subsection{Efficient Max-Pooling ($mp$)}

\begin{prob}
\label{prob:2}
Given a matrix $\llbracket \textbf{X} \rrbracket^{m \times m}$ which is element-wisely encrypted under the Paillier HE scheme (  $x_{ij}\in \mathbb{Z}_n^*$), the server computes $mp(\llbracket \textbf{X} \rrbracket^{m \times m}, t, s)$, without leaking the information of any element $x_{ij}$. $t$ is the pooling-window dimension and $s$ is the stride, where $s\leq t< m$ (usually, $t=s=2$). The client has $sk$, the server has $pk$ but cannot access to $sk$.
\end{prob}

The max-pooling ($mp$) operation outputs the maximum value of each pooling window (in size of $t \times t$). For succinctness, we denote the pooling window in form of vector, i.e., $max(\{ \llbracket x_0 \rrbracket,\llbracket x_1 \rrbracket,\cdots,\llbracket x_m \rrbracket \})$, $m=t\times t$. In existing solutions, e.g.,\cite{juvekar2018gazelle,riazi2019xonn,liu2017oblivious}, the $mp$ operation is notoriously inefficient. 

In this section, we will show how to efficiently compute the $mp$ layer. We first introduce a new protocol to compute $\max(\llbracket x_i \rrbracket, \llbracket x_j \rrbracket)$, which is the fundamental building block for computing the $mp$ layer. Then, we leverage the $\llbracket x_i \rrbracket \ominus \llbracket x_j \rrbracket$ inherent in the $\max(\llbracket x_i \rrbracket, \llbracket x_j \rrbracket)$ protocol to reduce homomorphic computations. Lastly, we present a method to absorb the $relu$ computation into the $mp$ layer.

\subsubsection{Compute $\max(\llbracket x_i \rrbracket, \llbracket x_j \rrbracket)$}
The $\max(\llbracket x_i \rrbracket, \llbracket x_j \rrbracket)$ is the fundamental computation for the max-pooling operation. We construct a lightweight protocol to compute $\max(\llbracket x_i \rrbracket, \llbracket x_j \rrbracket)$. The basic idea is first to covert it to a comparison between a ciphertext and 0 (i.e., $\max(\llbracket x\rrbracket,0 )$), then recover the result from the client's response, which is similar to the secure $relu$ computation protocol.  We summarize the $\max(\llbracket x_i \rrbracket, \llbracket x_j \rrbracket)$ computation protocol in Fig.~\ref{tab:cmp2cipher}.

\begin{figure}[h!]
\centering
\centering\framebox{
\begin{tabular}{llll}
%\hline
$\downarrow$&\textbf{Server}  ($(\llbracket x_i \rrbracket,\llbracket x_j \rrbracket), pk)$,                                                              &                                                                           &                                                                                \\ \hline
1&$\tau \sample  \mathbb{Z}_n^+ \ \  s.t. \ \ \gcd(\tau, n)=1$ &                                                                           &                                             \\
2&$\llbracket y  \rrbracket_s := (\llbracket x_i \rrbracket \ominus \llbracket x_j \rrbracket) \odot \tau$ &                                                                           &                                             \\
3& & $\xrightarrow{\llbracket y  \rrbracket_s}$ &                                                                                       \\ \hline
$\downarrow$& \textbf{Client} ($\llbracket y  \rrbracket_s, pk, sk$) &  &                                                                                       \\ \hline
                                                                 
5&  $\quad y_s :=$\emph{HDec}$(\llbracket y  \rrbracket_s,sk)$                                                               &                                                                           &  \\
6& $\quad y_c :=y_s$ if $y_s>0$ else $0$  &                                                                           & \\
7& $\quad \llbracket y  \rrbracket_c := $\emph{HEnc}$(y_c,pk)$                                                                                                                                &                                                                           &  \\       8&                                                                                                                          
                                                                 & $\xrightarrow{\llbracket y \rrbracket_c}$                             &                                                                                       \\ \hline
$\downarrow$&  \textbf{Server} ($\llbracket y \rrbracket_c, pk$)                                                                                                                       
                                                                 &                            &                                                                                       \\ \hline
9& $\llbracket y  \rrbracket_s := \llbracket x_j \rrbracket +\llbracket y \rrbracket_c \odot \tau^{-1}$  &                                                                           &                                                                                       \\
%\Comment{post-processing $\hat{\textbf{r}}_v$}   &                                                                           &                                                                                       %\\ \hline
\end{tabular}
}
\caption{Secure $\max(\llbracket x_i \rrbracket, \llbracket x_j \rrbracket)$ computation protocol. $\llbracket x_i \rrbracket, \llbracket x_j \rrbracket$ are randomly selected from a pooling window, such that we can sample the binder factor $\tau$ from $\mathbb{Z}^+$.}
\label{tab:cmp2cipher}
\end{figure}

For each pooling window, the max-pooling operation needs $m-1$ comparisons and calls the $\max(\llbracket x_i \rrbracket, \llbracket x_j \rrbracket)$ protocol $\left \lceil log_2(n) \right \rceil$ times. The same with the $relu$ protocol (see Fig.~\ref{tab:completedrelu}), we adopt the random shuffling method to avoid leaking $x_i - x_j  \stackrel{?}{=} 0$. Firstly, the server randomly maps the elements of each pooling window into pairs; then, the server shuffles all the pairs from all the pooling windows (each pair as an unit); the third, the server executes the secure $\max(\llbracket x_i \rrbracket, \llbracket x_j \rrbracket)$ computation protocol to get the maximum of each pair.

\subsubsection{$\llbracket x_i \rrbracket \ominus \llbracket x_j \rrbracket$ Benefits Computation} The secure $\max(\llbracket x_i \rrbracket, \llbracket x_j \rrbracket)$ computation protocol contains a subtraction between two ciphertexts (i.e., $\llbracket x_i \rrbracket \ominus \llbracket x_j \rrbracket$), we can exploit this fact to reduce computations on encrypted data.

\begin{figure}[h]
\centering
\includegraphics[scale=0.7]{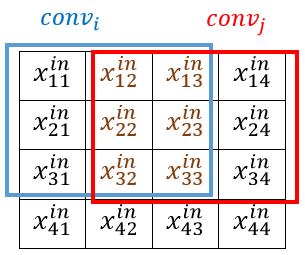}
\caption{An illustration of two adjacent $conv$ operations, with stride $s=1$ and $conv$ window size $w=3$. $ x_i = conv_i,  x_j=conv_j$. $x_*^{in}$ indicates the element of the $conv$ inputs.}
\label{fig:xixj}
\end{figure}

As shown in Fig.~\ref{fig:xixj}, $\llbracket x_i \rrbracket=conv_i= \sum_{u=1}^3\sum_{v=1}^3a_{ij}\llbracket x_{ij}^{in} \rrbracket$ and $\llbracket x_j \rrbracket=conv_j=\sum_{u=1}^3\sum_{v=2}^4a_{ij}\llbracket x_{ij}^{in} \rrbracket$ are the results of two adjacent $conv$ operations. Each $conv$ costs $w^2$ homomorphic multiplication-and-additions (here, $w=3$). Instead of independently computing $conv_i$ and $conv_j$, we leverage the computation of $\llbracket x_i \rrbracket \ominus \llbracket x_j \rrbracket$ to reduce HE operations as follows,
\begin{equation}
\label{eq:2conv}
   \begin{split}
    & \llbracket x_i \rrbracket \ominus \llbracket x_j \rrbracket = \sum_{u=1}^3\sum_{v=1}^3a_{ij} \llbracket x_{ij}^{in} \rrbracket \ominus \sum_{u=1}^3\sum_{v=2}^4a_{ij}\llbracket x_{ij}^{in} \rrbracket\\
    &= \sum_{u=1}^3\sum_{v=2}^3(a_{uv}-b_{uv})\llbracket x_{uv}^{in}\rrbracket \oplus \sum_u^3x_{u1}^{in} \ominus \sum_u^3 \llbracket x_{u4}^{in} \rrbracket
   \end{split}
\end{equation}
where $a_{uv}$ and $b_{uv}$ denote the weights of the two $conv$ filters, respectively. As shown in Equation~(\ref{eq:2conv}), for two adjacent $conv$ operations, we only need to apply one homomohphic multiplication-and-addition for each element of the overlapped region (i.e., by $(a_{uv}-b_{uv})\llbracket x_{uv}^{in}\rrbracket$).

In general, suppose the $conv$ window size is $w$ and the stride is $s$. Thanks to the $\llbracket x_i \rrbracket \ominus \llbracket x_j \rrbracket$, we can reduce the number of homomorphic multiplication-and-additions (of two $conv$ operations) from $2w^2$ to $w(w-s)+2ws=w^2+ws$. The ratio of computation reduction is $1-\frac{w^2+ws}{2w^2}=0.5-\frac{s}{2w}$. In most CNNs, $s\in \{1,2\}$ and $w\in \{3,5,7,11\}$. Therefore, nearly 50\% of homomorphic multiplication-and-additions can be reduced when meeting a $mp$ layer.

\subsubsection{Compute $relu\rightarrow mp$} Usually, a $mp$ layer often directly follows a $relu$ layer, i.e., $relu \rightarrow mp$~\cite{he2016deep,krizhevsky2012imagenet}. For each pooling window, we can write the computation of $relu \rightarrow mp$ as follows,
\begin{equation}
 \label{eq:rmax}
\max(\max(x_1,0),\max(x_2,0),\cdots,\max(x_m,0))
\end{equation}
Computing the Equation~(\ref{eq:rmax}) costs $2n-1$ comparisons. Clearly, we can transform this equation into following form,
 \begin{equation}
  \label{eq:maxrelu}
 \begin{split}
 &\max(\max(x_1,0),\max(x_2,0),\cdots,\max(x_m,0)) \\
=&\max(max(x_1,x_2,\cdots,x_m),0 )      
 \end{split}
\end{equation}
Compared with Equation~(\ref{eq:rmax}), Equation~(\ref{eq:maxrelu}) reduces the number of comparisons from $2m-1$ to $m$. From the perspective of communication overhead, by this transformation, we can compute the max-pooling layer for free.

 \subsubsection{Summary} Computing the $mp$ layer often leads to expensive communication overhead in existing methods~\cite{juvekar2018gazelle,liu2017oblivious}. A commonly-seen trick is to reduce the use of max-pooling layers such as using mean-pooling layers instead. However, this approach may results in a risk of accuracy degradation, as it breaches the original design.
 
 In this section, we first present a fast method for computing the max-pooling layer, of which the communicational overhead is equivalent to computing a $relu$ layer. Then, we exploit the stable design pattern in CNN, i.e., $conv \rightarrow relu \rightarrow mp$, to further reduce the computational cost, and to absorb the computation of $relu$ into $mp$ (looks like we can compute the $mp$ layer for free). It means that, in the Popcorn framework, the $mp$ layer can be a factor to improve efficiency, instead of becoming a heavy burden as usual.

%\section{Inference Acceleration via Network Compressing}
%-------------------------------------------------------------
\section{Popcorn: Fast Linear Computation}
%-------------------------------------------------------------

In the Popcorn framework, the input of each layer is element-wisely encrypted (the client data is the first layer input). Based on this fact, we can exploit neural network compression technologies to reduce HE computations in linear layers (i.e., the $conv$ and $fc$ layers), speeding up the oblivious inference. In this section, we introduce fast linear computation methods, which rely on pruned-and-quantized networks and binarized neural networks, respectively. Usually, the former preserves accuracy well; the latter leads to higher efficiency. %But it is not as accurate as the former, in particular, for those commercial-level neural networks, e.g., VGG on ImageNet~\cite{qin2020binary}.

%that fact that the client data is element-wisely encrypted, together with network pruning and quantization~\cite{han2015deep,yang2017designing}, to accelerate oblivious inference. 
\subsection{Pruning-Then-Quantization}
\label{subsec:ptq}

Suppose there are two filters (represented in a form of vector) $\textbf{a}= [a_1, a_2, \cdots, a_{n-1}, a_{n}]$, $\textbf{b}= [b_1, b_2, \cdots, b_{n-1}, b_{n}]$, and an encrypted input $\llbracket \textbf{x} \rrbracket= [\llbracket x_0 \rrbracket, \llbracket x_1 \rrbracket,\cdots, \llbracket x_{n-1} \rrbracket, \llbracket x_n \rrbracket]$. Before performing the $conv$ operations, we can adopt network pruning techniques to discover and remove un-important weights of $\textbf{a},\textbf{b}$ (e.g., let $a_2=0,\ a_{n-1}=0,\ b_1=0$), and employ network quantization methods to force multiple weights of $\textbf{a},\textbf{b}$ to share the same value (e.g., $a_1=b_1$). We illustrate how to use the prunned-and-quantized $conv$ filters to reduce homomorphic computations as follows,

\begin{equation}
\label{eq:dotp}
    \begin{split}
& \begin{vmatrix}
0 & a_1 & a_2 & \cdots & 0 &a_{n} \\ 
b_0 & b_1 & 0 & \cdots & b_{n-1} &b_{n}
\end{vmatrix}
\begin{vmatrix}
\llbracket x_0 \rrbracket \\ 
\llbracket x_1 \rrbracket \\ 
\llbracket x_2 \rrbracket \\ 
\cdots\\ 
\llbracket x_{n-1} \rrbracket \\ 
\llbracket x_{n} \rrbracket
\end{vmatrix} \\
=&\begin{vmatrix}
0+a_1\llbracket x_1 \rrbracket +a_2\llbracket x_2 \rrbracket+\cdots+0+a_n\llbracket x_n \rrbracket\\ 
b_0\llbracket x_0\rrbracket+0+0+\cdots+b_{n-1}\llbracket x_{n-1} \rrbracket+a_n\llbracket x_n \rrbracket
\end{vmatrix}
    \end{split}
\end{equation}

Firstly, we can simply skip any computations related to the removed weights (which are permanently set to 0). Then, we can find out which weights connect to the same ciphertext and share the same value, thus reusing the intermediate results. For example, if $a_i=b_i$ (where $0\leq i \leq n$) we can reuse the result of $a_i\llbracket x_i \rrbracket$ when computing $b_i\llbracket x_i \rrbracket$. Obviously, the inference acceleration relies on the number of removed weights (by pruning) and reused intermediate results (by quantization). This observation straightforwardly applies to the $fc$ layer.

\subsubsection{Network Compression For Popcorn}
A number of network compression methods have been proposed for different purposes such as minimizing model size~\cite{han2015deep}, reducing energy consumption~\cite{yang2017designing}. However, all the exiting methods are designed for plaintext inputs. There is a lack of investigation for network compression for ciphertext inputs. In this section, we analyze the weight pruning and quantization methods that are fit for the Popcorn framework.

\emph{Weight Pruning.} 
The aim of weight pruning is to further reduce the number of homomorphic computations, by removing more weights. For a $conv$ layer, each filter weight connects to  $\frac{w^2}{s^2}$ ciphertexts, where $w$ is the input dimension of the layer, $s$ is the stride-size of the $conv$ operations. So a single weight in different $conv$ layers may connect to a different number of ciphertexts. This means that removing the same number of weights from different $conv$ layers can result in a different reduction of homomorphic operations, depending on the input dimension $w$ of each $conv$ layer. The $fc$ layer is a vector-matrix multiplication, each weight connects to one ciphertext. Therefore, for $conv$ layers, it is suggested to first remove weights in a layer that has a larger input dimension.

\emph{Weight Quantization.} 
The purpose of weight quantization is to reuse the intermediate results computed between weights and encrypted inputs, as much as possible. Hence, the quantization priority of each layer depends on the number of weights that a ciphertext connects to, i.e., the larger the number, the higher the priority of the layer is. The weights of a high priority layer should first be quantized into lower bit representation. For a $conv$ layer, each ciphertext connects to $\frac{c_of_w^2}{s^2}$ weights, where $c_o$ is the number of filters and $f_w$ is the filter dimension. Assume the weights are represented in $t$ bits, the reuse ratio is $\geq \frac{c_ot^2}{s^2\cdot 2^m}$. For a $fc$ layer, each ciphertext connects to $c_o$ weights, here we  re-define $c_o$ as the $fc$ layer output dimension. The reuse ratio is $\geq \frac{c_o}{2^m}$.  With a quantized network, we can limit the number of homomorphic computations to $O(n\cdot 2^m)$, where we abuse $n$ to denote the number of ciphertexts. When the $c_0$ and $f_w$ are large while the $m$ and $s$ are small, the reduced homomorphic computations are significant.

\subsubsection{Summary}
\label{subsec:compressionsummary}

Based on the analysis above, we suggest a layer-by-layer pruning-then-quantization approach to speed up the linear layer computations in the Popcorn framework. This approach starts from pruning layers with the largest input dimensions or with the most number of multiplication-and-accumulation operations. It can be implemented through the algorithm introduced by~\cite{yang2017designing}. After the pruning process, we can quantify the remaining non-zero weights into low bits, beginning with the layers in which each ciphertext connects to the most non-zero weights. For a specific layer, we can adopt the codebook-based method introduced in~\cite{han2015deep}. Compared with other methods, this method can obtain a lower-bit representation while ensuring the same accuracy.

\subsection{Binarized Neural Network}
\label{subsec:bnn}

The binarized neural network is an extreme case of network quantization, of which the weights are binarized, i.e., $\{-1,+1 \}$. In this section, we introduce how to efficiently evaluate a binarized network in the Popcorn framework.

Assume there are two binarized filters (in the form of vector) $\textbf{a}= [+1, -1, \cdots, -1, +1]$, $\textbf{b}= [-1, -1, \cdots, +1, -1]$, and an encrypted input $\llbracket \textbf{x} \rrbracket= [\llbracket x_0 \rrbracket, \llbracket x_1 \rrbracket,\cdots, \llbracket x_{n-1} \rrbracket, \llbracket x_n \rrbracket]$. We describe the $conv$ (as well as the $fc$) computation as follows, 

\begin{equation}
\label{eq:dotp}
    \begin{split}
& \begin{vmatrix}
+1 & -1  & \cdots & -1 &+1 \\ 
-1 & +1  & \cdots & -1 &-1
\end{vmatrix}
\begin{vmatrix}
\llbracket x_0 \rrbracket \\ 
\llbracket x_1 \rrbracket \\ 
\cdots\\ 
\llbracket x_{n-1} \rrbracket \\ 
\llbracket x_{n} \rrbracket
\end{vmatrix} \\
=&\begin{vmatrix}
+\llbracket x_0 \rrbracket +\llbracket x_1 \rrbracket+\cdots+\llbracket x_{n-1} \rrbracket-\llbracket x_n \rrbracket\\ 
-\llbracket x_0 \rrbracket -\llbracket x_1 \rrbracket+\cdots+\llbracket x_{n-1} \rrbracket-\llbracket x_n \rrbracket
\end{vmatrix}
    \end{split}
\end{equation}

It is clear that the computations only rely on efficient homomorphic additions, avoiding expensive multiplications. Therefore, the execution of $conv$ and $fc$ layers can be very efficient. In the sub-section~\ref{subsec:plus1}, we introduce the "$+1$ Trick" to further improve efficiency.

\subsubsection{$+1$ Trick}
\label{subsec:plus1}
We exploit the fact that the weights are binarized, $\textbf{a} \in \{+1,-1\}^m$, to roughly halve the computation cost, through a $+1$ trick as follows, 
\begin{equation}
\label{eq:plusone}
\begin{split}
    \textbf{a}\textbf{x} & =  (\textbf{1}+\textbf{a})\textbf{x}-\textbf{1}\textbf{x}\\
    & =\frac{(\textbf{1}+\textbf{a})}{2}\textbf{x}+\frac{(\textbf{1}+\textbf{a})}{2}\textbf{x} -\textbf{1}\textbf{x}
\end{split}
\end{equation}
where $\textbf{1}$ indicates a vector in which each element is 1. Two avoid multiplications, we split $(\textbf{1}+\textbf{a}) \in \{0, +2 \}^m$ into two pieces of $\frac{(\textbf{1}+\textbf{a})}{2} \in \{0, +1 \}^m$. For the same layer, we only need to compute $\textbf{1}\textbf{x}$ once, and the cost can be amortized by all the $conv$ filters. For the case that there are more $+1$ than $-1$, we can adapt the Equation~(\ref{eq:plusone}) to $\textbf{a}\textbf{x}=-\frac{(\textbf{1}-\textbf{a})}{2}\textbf{x}-\frac{(\textbf{1}-\textbf{a})}{2}\textbf{x} +\textbf{1}\textbf{x}$. Therefore, we can always halve the homomorphic computations by the $+1$ trick. Different from~\cite{sanyal2018tapas}, we don't need to binarize the input $\textbf{x}$, facilitating accuracy preservation.

\section{Comparison and New Benchmark}
In this section, we first conduct a general comparison between the Popcorn and prior arts regarding the privacy guarantee and utility (Section~\ref{subsec:gcopa}); then we test Popcorn and compare it with previous arts in term of efficiency; Lastly, we report the benchmarks of oblivious inference on the ImageNet dataset, based on start-of-the-art networks.

\subsection{Evaluation Settings}
We implement the Popcorn framework based on OPHELib ~\cite{ophelib}, which provides an implementation of the Paillier encryption scheme, written in C++. The code is compiled using GCC with the '-O3' optimization, and the OpenMP for parallel acceleration is activated.  The test
is performed on (Ubuntu 18.04 LTS) machines with Intel(R) Xeon(R) CPU E5 and 32GB of RAM. The Paillier key size is always set to 2048 bits. We execute the comparison and benchmarks on the MNIST, CIFAR-10, and the ILSVRC2012 ImageNet dataset. To the best of our knowledge, this is the first report for benchmarking oblivious inference on the ImageNet dataset.

\subsection{General Comparison of Prior Arts}
\label{subsec:gcopa}
Existing frameworks are usually efficiency-oriented, with different compromises in terms of privacy and computational guarantees. For clarification, we describe prior arts and our Popcorn framework according to the following guarantees,
\begin{itemize}
    \item \emph{P1 (data privacy).} The framework hides the client data from the server, except for data dimensions.
    \item \emph{P2 (network privacy).} The framework hides the server's network weights (as well the output values of each hidden layer) from the client.
    \item \emph{P3 (network privacy).} The framework hides the server's network (including network weights, architecture and the output values of each hidden layer) from the client, except for (1) the number of layers; (2) the number of activations of each layer; (3) classification results.
    \item \emph{P4 (network privacy).} The framework hides the server's network from the client, except for classification results.
    \item \emph{V1.} The framework supports any type of CNNs.
    \item \emph{V2.} The framework does not rely on any external party.
\end{itemize}

\begin{table}[h]
\centering
\begin{tabular}{l|l|l|l|l|l|l}
\hline
Framework   & P1 & P2 & P3 & P4 & V1 & V2 \\ \hline
CryptoNets          & $\checkmark$   & $\checkmark$  & $\checkmark$  & $\checkmark$    & $-$ & $\checkmark$ \\ \hline
SecureML       & $\checkmark$   &  $\checkmark$  & $-$ &  $-$   & $-$ & $-$ \\ \hline
MiniONN       & $\checkmark$   & $\checkmark$  & $-$  &  $-$   & $\checkmark$ & $\checkmark$ \\ \hline
Gazelle        & $\checkmark$   & $\checkmark$  & $\checkmark$  & $-$   & $\checkmark$ & $\checkmark$ \\ \hline
XONN              & $\checkmark$   & $\checkmark$  & $-$   & $-$   & $-$ & $\checkmark$  \\ \hline
\textbf{Popcorn}             & $\checkmark$   & $\checkmark$  & $\checkmark$ &  $-$   & $\checkmark$ & $\checkmark$ \\ \hline
\end{tabular}
\caption{The comparison of different frameworks.}
\label{tab:frameworkcompare}
\end{table}

As shown in Table~\ref{tab:frameworkcompare}, all the frameworks can meet the $P1$ criteria, i.e., the client data privacy is well preserved. However, the network information is leaked to different extents within different frameworks. CryptoNets~\cite{gilad2016cryptonets} preserves the most network privacy ($P2,P3,P4$, assume the encryption parameters are large enough), while SecureML~\cite{mohassel2017secureml}, MiniONN~\cite{liu2017oblivious} and XONN~\cite{riazi2019xonn} leak the most network information (i.e., $P3, P4$ are leaked). Gazelle~\cite{juvekar2018gazelle} and Popcorn achieve a compromise ($P2, P3$ are protected). Note that, to improve efficiency, Gazelle introduced a padded patch, but pay the price of privacy, i.e., the $conv$ filter size is disclosed to the client. So far, it is very challenging to completely protect network privacy. For example, in CryptoNets, the activation method and network size can be inferred by the encryption parameters; in Gazelle and Popcorn, the size of each network layer can be deduced by the number of neurons. Though multiparty-computation-based frameworks do not completely protect network privacy, we believe that it is important to hide more information, rather than ignore the privacy breach or directly disclosing the network information to the client.

The Versatility, i.e., the support for various CNNs ($V1$) and the requirements of server settings ($V2$), directly impact the deployments in reality. CryptoNets and SecureML only support linearized CNNs,  which require substituting non-linear activation functions with polynomials. This approach may significantly reduce accuracy performance, especially for large neural networks.  The SecureML needs two non-colluding servers, which may narrow the applicable scenarios. XONN is applicable only when the weights and activations of a network are binarized. Gazelle, MiniONN, and Popcorn satisfy both the $V1$ and $V2$ criteria, as they don't need to adjust the original design of a network. It is worth mentioning that, using XONN, the client will undertake the most computational overhead, as it is responsible for executing the complied network circuits to obtain classification results. It may become a heavy burden to the client when the network becomes large.

\subsection{Efficiency Comparison with Prior Arts}
%1. why mnist cifar10
To compare with prior arts, we report runtime (RT), communication bandwidth (COM) and accuracy (acc\%) on MNIST and CIFAR-10 classification tasks. Since the Popcorn can leverage compressed and binarized networks to accelerate the oblivious inference. We implement two versions of the Popcorn framework, Popcorn-b and Popcorn-c. The former supports binarized neural networks (section~\ref{subsec:ptq}) and the latter supports compressed neural networks ( section~\ref{subsec:bnn}).

For network binarization, we follow the XNOR-Net method~\cite{rastegari2016xnor} that we only binarize the network weights and leave the activation values as the original. By this, we can preserve the information carried on activations, benefiting accuracy performance~\cite{qin2020binary}. For network compression, we adopt the layer-by-layer pruning-then-quantization approach suggested in Section~\ref{subsec:compressionsummary}. Firstly, we adopt the layer-by-layer weight pruning method introduced by~\cite{yang2017designing}, to remove more weights in lower layers. Then, we employ the deep-compression method~\cite{han2015deep} to quantize the rest non-zero weights, to use lower bits to represent weights in $fc$ layers, and $conv$ layers which contain more filters. 

As mentioned above, in the implementation, we build the network binarization and the pruning-then-quantization approaches with existing arts (i.e.,~\cite{rastegari2016xnor, yang2017designing,han2015deep}), thus avoiding tedious accuracy performance evaluation and focusing on the efficiency testing and comparison.

\subsubsection{Evaluation on MNIST} 

The MNIST is an entry-level image classification dataset. It consists of a set of grayscale images of handwritten digits (i.e., [0,9]), and the dimension of each image is $28\times 28 \times 1$. We perform the experiments with three classical neural networks which were frequently adopted by previous arts, as shown in Table ~\ref{tab:netonmnists}. 
\begin{table}[h]
\centering
\begin{tabular}{l|l|l}
\hline
Network & Source                                                                         & Decription         \\ \hline
NM1     & \begin{tabular}[c]{@{}l@{}}CryptoNets, XONN\\ Gazelle, MiniONN\end{tabular}    & 3 $fc$               \\ \hline
NM2     & \begin{tabular}[c]{@{}l@{}}CryptoNets, XONN\\ DeepSecure, Gazelle\end{tabular} & 1 $conv$, 2 $fc$       \\ \hline
NM3     & \begin{tabular}[c]{@{}l@{}}XONN, Gazelle,\\ MiniONN\end{tabular}               & 2 $conv$, 2 $mp$, 2 $fc$ \\ \hline
\end{tabular}
\caption{Network architectures for the MNIST dataset. NM stands for \textbf{N}etwork on \textbf{M}NIST. NM1 is a MLP network. NM2 and NM3 are two small CNNs. In CryptoNets, the $relu$ activation layer is replaced by the square ($f(x)=x^2$) activation layer. For detailed architecture information, please refer to the papers listed in the "Source" column.}
\label{tab:netonmnists}
\end{table}

The "P/Q" presents the pruning ratio (i.e., the number of removed weights divided by the number of total weights) and the average number of bits to represent the overall weights. As shown in Table~\ref{tab:mnist}, we can remove more than 90\% weights and quantize the rest weights to $\leq 6$ bits without losing accuracy. Sometimes, the Popcorn-c even results in higher accuracy.

\begin{table}[h]
\centering
\begin{tabular}{l|l|l|l|l|l}
\hline
   &P/Q     &    Framework      & RT    & COM       & acc\%          \\ \hline
    &     & MiniONN            & 1.04           & 15.8           & 97.6           \\ \cline{3-6} 
NM1 & 0.92/6    & Gazelle            & 0.09           & 0.5            & 97.6           \\ \cline{3-6} 
    &     & XONN               & 0.13           & 4.29           & 97.6           \\ \cline{3-6} 
    &     & \textbf{Popcorn-b} & \textbf{0.30}  & \textbf{0.78}  & \textbf{97.6}  \\ \cline{3-6} 
    &     & \textbf{Popcorn-c} & \textbf{0.51}  & \textbf{0.78}  & \textbf{98.4}  \\ \hline
    &     & CryptoNets         & 297.5          & 272.2          & 98.95          \\ \cline{3-6} 
    &     & DeepSecure         & 9.67           & 791.0          & 98.95          \\ \cline{3-6} 
    &     & MiniONN            & 1.28           & 47.6           & 98.95          \\ \cline{3-6} 
NM2 & 0.91/5.1    & Gazelle            & 0.29           & 8.0            & 99.0           \\ \cline{3-6} 
    &     & XONN               & 0.16           & 38.28          & 98.64          \\ \cline{3-6} 
    &     & \textbf{Popcorn-b} & \textbf{0.60}  & \textbf{1.90}   & \textbf{98.61} \\ \cline{3-6} 
    &     & \textbf{Popcorn-c} & \textbf{0.67}      & \textbf{0.84}      & \textbf{98.91}      \\ \hline
    &     & MiniONN            & 9.32           & 657.5          & 99.0           \\ \cline{3-6} 
    &     & Gazelle            & 1.16           & 70.0             & 99.0           \\ \cline{3-6} 
NM3 &  0.90/5.6   & XONN               & 0.15           & 62.77          & 99.0           \\ \cline{3-6} 
    &     & \textbf{Popcorn-b} & \textbf{7.14} & \textbf{19.50} & \textbf{99.0}  \\ \cline{3-6} 
    &     & \textbf{Popcorn-c} & \textbf{5.71}   & \textbf{10.50}  & \textbf{98.8}  \\ \hline
\end{tabular}
\caption{Comparison on MNIST. RT means the runtime in second. COM presents the communication bandwidth in megabyte. }
\label{tab:mnist}
\end{table}

For the binarized model evaluation, we did not adopt the scaling-factor method introduced by XONN~\cite{riazi2019xonn}. Instead, for all the networks, first we double the output size of the first layer, the others remain the same; then, we follow the XNOR-Net training method~\cite{rastegari2016xnor} to obtain binarized models. There are two main reasons leading us to the current implementation. First, we don't quantize the activation values, more information can be carried to preserve accuracy performance. Second, the networks for the MNIST classification tasks are very small (e.g., small input dimensions, $28\times 28 \times 1$). Instead of using a complicated method to discover a thin network architecture, we can empirically and efficiently try different settings of the neural networks. 

Regarding the runtime performance, XONN and Gazelle are in the leading position, the Popcorn follows. For the communication overhead, the Popcorn framework shows a significant advantage.  According to the results, all the frameworks achieve equivalent accuracy (from 97.6\% to 99.0\%, see Table~\ref{tab:mnist}). XONN and CryptoNets require modifying the original network design to fit the constraints of the adopted crypto primitives.  The CryptoNets shows that by substituting the non-linear activation method (e.g., $relu$) with a square function, it can still get decent accuracy; XONN demonstrates that it is possible to improve accuracy by scaling up the network architecture size, remedying the accuracy loss caused by the binarizing network weights and activations. However, a natural question arises, can such tricks be applied to larger datasets (e.g., larger input dimensions and network size)?

\subsubsection{Evaluation on CIFAR-10} 
The CIFAR-10 is another popular image classification dataset, which consists of a number of colorful images and categorized into 10 classes such as bird, truck, cats. The dimension of each image is $32\times 32 \times 3$. Unlike the MNIST, the CIFAR-10 classification tasks require sophisticated design on neural networks. The tricks applied to the MNIST for preserving accuracy may not work on the CIFAR-10. For example,  Li et at.~\cite{chou2018faster} investigated different approximation methods to replace non-linear activation functions, but none of them could obtain decent accuracy. Straightforwardly increasing architecture size also becomes struggling to improve accuracy (see Table~\ref{tab:cifar10acc}). We conduct the experiments with two networks adopted by prior arts and a VGG variant for CIFAR-10~\cite{simonyan2014very,zhang2018lq}, summarized in Table~\ref{tab:networkonciarf10}.

\begin{table}[h]
\centering
\begin{tabular}{l|l|l}
\hline
Network & Source                                                           & Description         \\ \hline
NC1     & \begin{tabular}[c]{@{}l@{}}XONN, Gazelle,\\ MiniONN\end{tabular} & 9 $conv$, 3 $mp$, 1 $fc$  \\ \hline
NC2     & XONN                                                             & 10 $conv$, 3 $mp$, 1 $fc$ \\ \hline
VGG-c     & \cite{zhang2018lq}                                                             & 6 $conv$, 5 $mp$, 1 $fc$ \\ \hline
\end{tabular}
\caption{Networks on CIFAR-10. NC stands for \textbf{N}networks on \textbf{C}IFAR-10. VGG-c is a customized network for CIFAR-10~\cite{zhang2018lq}. For detailed architecture information, please refer to the papers listed in the "Source" column.}
\label{tab:networkonciarf10}
\end{table}

For network binarization, we also use the XNOR-Net training method. For the NC1 and NC2, we double the number of $conv$ filters of the first 3 layers to obtain equivalent accuracy with their full precision version. For VGG-c, we retain the original network architecture. Note that VGG-c has many more filters than NC1 and NC2. In Popcorn-b, we don't binarize activations. As shown in Table~\ref{tab:cifar10acc}, the network NC1 leads to a 10\% decrease in accuracy. Therefore, simply scaling up the size of a fully binarized network may not obtain decent accuracy as expected, instead, it is suggested to use state-of-the-art networks for classification tasks.

\begin{table}[h]
\centering
\begin{tabular}{l|l|l|l|l|l}
\hline
             &P/Q          &  Framework          & RT & COM & acc\% \\ \hline
            &         & MiniONN    & 544.0       & 9272.0    & 81.61 \\ \cline{3-6} 
            &         & Gazelle    & 15.48       & 1236.0    & 81.61 \\ \cline{3-6} 
NC1         &  0.76/6.7       & XONN       & 5.79        & 2599.0    & 81.61 \\ \cline{3-6} 
            &         & \textbf{Popcorn-b}  & \textbf{59.7}        & \textbf{125.5}     & \textbf{81.66} \\ \cline{3-6} 
            &         & \textbf{Popcorn-c}  & \textbf{84.5}            & \textbf{78.4}          & \textbf{81.74}      \\ \hline
NC2         &         & XONN       & 123.9       & 42362   & 88.0  \\ \cline{3-6} 
            &    0.77/7.1     & \textbf{Popcorn-b}  & \textbf{268.2}       & \textbf{704.7}     & \textbf{88.0}  \\ \cline{3-6} 
            &         & \textbf{Popcorn-c}  & \textbf{528.8}       & \textbf{207.8}     & \textbf{87.6}  \\ \hline
VGG-c       &  0.81/6.4       & \textbf{Popcorn-b} & \textbf{513.3}       & \textbf{449.8}     & \textbf{91.0}  \\ \cline{3-6} 
            &         & \textbf{Popcorn-c}  & \textbf{918.1}            & \textbf{449.8}     & \textbf{91.7}  \\ \hline
\end{tabular}
\caption{Comparison on CIFAR-10. RT means the runtime in second. COM presents the communication bandwidth in megabyte.}
\label{tab:cifar10acc}
\end{table}

As shown in Table~\ref{tab:cifar10acc}, at the same accuracy level, Popcorn requires much less communication overhead. For example, to reach the accuracy of 88.0\%, the communication overhead of XONN is around 41 GB, while Popcorn-b only needs 704.7 MB, which is $60 \times$ smaller. In addition, we use the state-of-the-art binarized VGG-c, getting an accuracy of 91\%, and communication bandwidth is only 450 MB. This observation shows again the importance of using start-of-the-art networks.

%Therefore, it is necessary to maintain a standard CNN architecture (i.e., CONV with non-linear activations). MiniONN, Gazelle and our Popcorn don't need to change the existing design, which are able to well preserve the accuracy. XONN requires binarized neual network, it tries to maintain accuracy by enlarging network size.

\subsection{Benchmarks on ImageNet}
%. why imagenet 
With the promising results on the datasets MNIST and CIFAR-10 (especially the communication overhead), we look at a commercial-level dataset, the ImangeNet ILSVRC2012 dataset. Neural networks for classification on this dataset often have large input dimensions (i.e., $224 \times 224 \times 3$), which is much larger than the input dimension adopted for MNIST ( $28 \times 28 \times 1$) and CIFAR-10 ( $32 \times 32 \times 3$). To the best of our knowledge, this is the first report for benchmarking the oblivious inference on ImageNet classification tasks.

To evaluate the Popcorn framework, we adopt AlexNet~\cite{krizhevsky2012imagenet} and VGG~\cite{simonyan2014very}, which are two milestone networks for the ImageNet classification tasks, to benchmark the oblivious inference. Compared with the networks used for MNIST and CIFAR-10 classifications, the most significant difference is that the dimensions of input and each hidden layer are much larger. For ease of future comparison, we describe the network architectures in Table~\ref{tab:archaleximgnet}  and Table~\ref{tab:archvggimgnet}, respectively.

\begin{table}[h]
\centering
\begin{tabular}{l|l|l|l|l|l}
\hline
Layer     & Input                     & Kernel & pw & s &  P/Q\\ \hline
$conv$1  & $224, 224, 3$  & 64     & 11     & 4     & 0.83/8    \\ \hline
$conv$2   & $55, 55,64$  & 192    & 5      & 1   & 0.92/7    \\ \hline
$mp$      & $55, 55, 192$ &   $-$     & 2      & 2      &        \\ \hline
$conv$3  & $27, 27, 192$  & 384    & 3      & 1   & 0.91/7     \\ \hline
$mp$     & $27, 27, 384$  &    $-$    & 2      & 2       &        \\ \hline
$conv$4  & $13, 13, 384 $  & 256    & 3      & 1     &  0.81/7    \\ \hline
$conv$5  & $13, 13, 256 $  & 256    & 3      & 1     &  0.74/7    \\ \hline
$mp$      & $13, 13, 256 $  &   $-$       & 2      & 2     &          \\ \hline
$fc$1   & $9216$         &    $ 4096 $      &   $-$       &   $-$     &  0.92/6      \\ \hline
$fc$2     & $ 4096 $         &   $ 4096 $        &   $-$       &  $-$    & 0.91/6      \\ \hline
$fc$3    & $ 4096$         &    $ 1000 $      &   $-$       &  $-$    &  0.78/6      \\ \hline
\end{tabular}
\caption{AlexNet~\cite{krizhevsky2012imagenet}: pw stands for the dimension of the max-pooling window; s is the stride. P/Q records the pruning ratio and the number of bits for weights representation of each layer.}
\label{tab:archaleximgnet}
\end{table}

\begin{table}[h]
\centering
\begin{tabular}{l|l|l|l|l|l}
\hline
Layer  & Input                     & Kernel & pw & s &         P/Q \\ \hline
$conv$1  & $224, 224, 3$  & 64    & 3      & 1                 & 0.53/8      \\ \hline
$mp$     &      $224, 224, 64$                     &  $-$      & 2      & 2      &   $-$    \\ \hline
$conv$2  & $112, 112, 64$   & 128    & 3      & 1                 & 0.71/8      \\ \hline
$mp$     &  $112, 112, 128$                        &   $-$     & 2      & 2      &     \\ \hline
$conv$3 & $56, 56, 128$  & 256    & 3      & 1                  & 0.67/7      \\ \hline
$conv$4 & $56, 56, 256$  & 256    & 3      & 1                  & 0.65/7      \\ \hline
$mp$     & $56, 56, 256$ &  $-$      & 2      & 2      &    $-$   \\ \hline
$conv$5  & $28, 28, 256$  & 512    & 3      & 1                 &  0.69/7     \\ \hline
$conv$6  & $28 , 28, 512$ & 512    & 3      & 1                & 0.73/7      \\ \hline
$mp$     & $28 , 28, 512$  &   $-$     & 2      & 2      & $-$      \\ \hline
$conv$7  & $14, 14, 512$   & 512    & 3      & 1                &  0.78/6     \\ \hline
$conv$8  & $14 , 14 , 512$ & 512    & 3      & 1                &  0.81/6      \\ \hline
$mp$     &  $14 , 14 , 512$ &    $-$    & 2      & 2      & $-$      \\ \hline
$fc$1 & $25088$     &  $4096$  &   $-$    &     $-$                  & 0.96/6      \\ \hline
$fc$2 & $4096$    &  $4096$   &  $-$     &    $-$                    & 0.96/6       \\ \hline
$fc$3     & $4096$         &  $1000$       &  $-$      &         $-$    & 0.79/6      \\ \hline
\end{tabular}
\caption{VGG~\cite{simonyan2014very}. For $fc$ layers, the output dimension is recorded in the "Kernel" column. }
\label{tab:archvggimgnet}
\end{table}

For network compression and binarization, we use the same method applied to the MNIST and CIFAR-10 classification tasks. Thanks to \cite{yang2017designing}, a pruned AlexNet model already exists, and it was pruned starting from lower layers. So we directly use it as the basis, and quantize the remaining non-zero weights through the deep compression method~\cite{han2015deep}. The compression results of each layer of AlexNet and VGG are summarized in the Table~\ref{tab:archaleximgnet}  and Table~\ref{tab:archvggimgnet}.

\begin{table}[h]
\centering
\begin{tabular}{|l|l|l|l|l|}
\hline
                         &  Framework         & RT(m) & COM & acc\% \\ \hline
\multirow{2}{*}{AlexNet} & Popcorn-b & 9.94  & 560.6  & 78.1  \\ \cline{2-5} 
                         & Popcorn-c & 29.9  & 560.6  & 79.3  \\ \hline
\multirow{2}{*}{VGG}     & Popcorn-b & 115.4 & 7391.5 & 87.2  \\ \cline{2-5} 
                         & Popcorn-c & 568.1 & 7391.5 & 90.1  \\ \hline
\end{tabular}
\caption{benchmarks on ImageNet: RT (m) means the runtime in minute; COM presents the communication bandwidth in megabyte. (\textcolor{red}{Note: this version corrects a naive but significant typo in our previous version. Previously, we mistakenly indicated the communication cost in COM(g), i.e., communication bandwidth in gigabyte. In fact, it should be COM as in this version, and COM presents the communication bandwidth in megabyte. It is megabyte, not gigabyte.})}
\label{tab:benchmarkonimgnet}
\end{table}
%COM(g) indicates communication bandwidth in gigabyte.
We report the runtime  and communication bandwidth in Table~\ref{tab:benchmarkonimgnet}. Compared with MNIST and CIFAR-10, the runtime is significantly increased, as the network size is orders of magnitude larger than that for MNIST and CIFAR-10. The communication bandwidth is still in a reasonable range, even lower than the bandwidth required by the prior arts to run the CIFAR-10 classification tasks. It is worth stressing that the bandwidth complexity of Popcorn is $O(n)$, where $n$ is the number of activations. The bandwidth complexity of XONN and Gazelle relies on the network architecture. For example, using XONN, we need to compile the whole network into circuits; using Gazelle, large input dimensions require high degree polynomials (the efficiency is also obviously affected). When we try to evaluate AlexNet using Gazelle, on the same machine that the Popcorn runs, the out-of-memory error always occurs (using the implementation provided by~\cite{juvekar2018gazelle}). An estimation of executing AlexNet with Gazelle, the communication overhead is at least 50 GB, the XONN is even worse. Therefore, the Popcorn can have a significant advantage for evaluating state-of-the-art networks (e.g., AlexNet and VGG).

%\subsection{Discussion}

\section{Related Work}
Barni et al.~\cite{barni2006privacy} initiated one of the earliest attempts for oblivious inference, using homomorphic encryption (HE). Since HE is not compatible with non-linear algebraic operations (e.g., comparison, division), they introduced a multiplicative obfuscation to hide intermediate computation results. However, this method leaks information about neural network weights~\cite{orlandi2007oblivious}. Gilad-Bachrach et al.~\cite{gilad2016cryptonets} replaced the non-linear activation function (e.g., $relu(0,x)$~\cite{nair2010rectified}) with a low-degree polynomial ($f(x)=x^2$), making the neural network fully compatible with HE. Several works~\cite{chou2018faster,brutzkus2019low,bourse2018fast} developed different methods to improve the CryptoNets paradigm, in terms of efficiency and accuracy. However, compared with other approaches, the results are still not promising.

Rouhani et al.~\cite{rouhani2018deepsecure} proposed garbled circuits based framework for the oblivious inference, where the server compiles a pre-trained network into circuits and sends the circuits to a client, the client gets the prediction results by evaluating the circuits. However, performing multiplication in GC has quadratic computation and communication complexity with respect to the bit-length of the input operands. This fact rises up a serious efficiency problem for a precise inference (which often needs a high bit-length of the input operands).  Riazi et al.~\cite{riazi2019xonn} leveraged fully binarized neural networks, of which the weights and activations are binary (i.e., $\{+1,-1\}$), to accelerate the GC-based oblivious inference. However, fully binarized neural networks are not stable in accuracy performance, especially, when used for classification tasks on large-scale datasets (e.g., ImageNet~\cite{krizhevsky2012imagenet}).

Liu et al.~\cite{liu2017oblivious} combined HE, GC and SPDZ~\cite{damgaard2012multiparty} to speedup the oblivious inference. It showed a hybrid approach can be promising in both efficiency and accuracy. To balance the communication overhead and accuracy, Mishra et al.~\cite{mishra2020delphi} proposed to replace partial non-linear activations with low-degree polynomials, the inference computation is similar to~\cite{liu2017oblivious}. However, the SPDZ-based methods directly reveal the network architecture to the client. Juvekar et al.~ \cite{juvekar2018gazelle} leveraged SIMD to accelerate the linear computation in the inference by packing multiple messages into one ciphertext, and they used GC to compute the $relu$ activation and max-pooling layers. Zhang et al.~\cite{zhang2021gala} improved this solution by reducing the permutation operations when performing dot-product based on packed ciphertexts. 

The trusted execution environment technology (TEE, e.g., Intel SGX~\cite{costan2016intel}) is also an interesting approach to build oblivious inference frameworks (e.g.,~\cite{tramer2019slalom}). Generally, most TEE-based frameworks are more efficient than cryptography-based solutions~\cite{tramer2019slalom}. However, this approach requires trust in hardware vendors and the implementation of the enclave. We leave this discussion of TEE-based solutions out of the scope of this paper.

\section{Conclusion}
This work presented a concise oblivious neural network inference framework, the Popcorn. This framework was completely built on the Paillier HE scheme. It is easy to implement, one only needs to replace the algebraic operations of existing networks with their corresponding homomorphic operations. We conducted experiments on different datasets (MNIST, CIFAR-10 and ImageNet), and showed its significant advantage in communication bandwidth. To our best knowledge, this work is the first report for oblivious inference benchmark on ImageNet-scale classification tasks.

%-------------------------------------------------------------------------------
%\section*{Acknowledgments}
%-------------------------------------------------------------------------------

%The USENIX latex style is old and very tired, which is why
%there's no \textbackslash{}acks command for you to use when
%acknowledging. Sorry.

%%-------------------------------------------------------------------------------
%\section*{Availability}
%%-------------------------------------------------------------------------------

%USENIX program committees give extra points to submissions that are
%backed by artifacts that are publicly available. If you made your code
%or data available, it's worth mentioning this fact in a dedicated
%section.

%-------------------------------------------------------------------------------
\bibliographystyle{plain}
%\bibliography{\jobname}
%\bibliographystyle{IEEEtran}
\bibliography{main.bib}

\begin{thebibliography}{10}

\bibitem{barni2006privacy}
Mauro Barni, Claudio Orlandi, and Alessandro Piva.
\newblock A privacy-preserving protocol for neural-network-based computation.
\newblock In {\em Proceedings of the 8th workshop on Multimedia and security},
  pages 146--151, 2006.

\bibitem{beimel2011secret}
Amos Beimel.
\newblock Secret-sharing schemes: a survey.
\newblock In {\em International conference on coding and cryptology}, pages
  11--46. Springer, 2011.

\bibitem{bellare2012foundations}
Mihir Bellare, Viet~Tung Hoang, and Phillip Rogaway.
\newblock Foundations of garbled circuits.
\newblock In {\em Proceedings of the 2012 ACM conference on Computer and
  communications security}, pages 784--796, 2012.

\bibitem{bourse2018fast}
Florian Bourse, Michele Minelli, Matthias Minihold, and Pascal Paillier.
\newblock Fast homomorphic evaluation of deep discretized neural networks.
\newblock In {\em Annual International Cryptology Conference}, pages 483--512.
  Springer, 2018.

\bibitem{brutzkus2019low}
Alon Brutzkus, Ran Gilad-Bachrach, and Oren Elisha.
\newblock Low latency privacy preserving inference.
\newblock In {\em International Conference on Machine Learning}, pages
  812--821. PMLR, 2019.

\bibitem{chou2018faster}
Edward Chou, Josh Beal, Daniel Levy, Serena Yeung, Albert Haque, and
  Li~Fei-Fei.
\newblock Faster cryptonets: Leveraging sparsity for real-world encrypted
  inference.
\newblock {\em arXiv preprint arXiv:1811.09953}, 2018.

\bibitem{costan2016intel}
Victor Costan and Srinivas Devadas.
\newblock Intel sgx explained.
\newblock {\em IACR Cryptol. ePrint Arch.}, 2016(86):1--118, 2016.

\bibitem{damgaard2012multiparty}
Ivan Damg{\aa}rd, Valerio Pastro, Nigel Smart, and Sarah Zakarias.
\newblock Multiparty computation from somewhat homomorphic encryption.
\newblock In {\em Annual Cryptology Conference}, pages 643--662. Springer,
  2012.

\bibitem{esteva2017dermatologist}
Andre Esteva, Brett Kuprel, Roberto~A Novoa, Justin Ko, Susan~M Swetter,
  Helen~M Blau, and Sebastian Thrun.
\newblock Dermatologist-level classification of skin cancer with deep neural
  networks.
\newblock {\em nature}, 542(7639):115--118, 2017.

\bibitem{gentry2009fully}
Craig Gentry et~al.
\newblock {\em A fully homomorphic encryption scheme}, volume~20.
\newblock Stanford university Stanford, 2009.

\bibitem{gilad2016cryptonets}
Ran Gilad-Bachrach, Nathan Dowlin, Kim Laine, Kristin Lauter, Michael Naehrig,
  and John Wernsing.
\newblock Cryptonets: Applying neural networks to encrypted data with high
  throughput and accuracy.
\newblock In {\em International Conference on Machine Learning}, pages
  201--210. PMLR, 2016.

\bibitem{han2015deep}
Song Han, Huizi Mao, and William~J Dally.
\newblock Deep compression: Compressing deep neural networks with pruning,
  trained quantization and huffman coding.
\newblock {\em arXiv preprint arXiv:1510.00149}, 2015.

\bibitem{he2016deep}
Kaiming He, Xiangyu Zhang, Shaoqing Ren, and Jian Sun.
\newblock Deep residual learning for image recognition.
\newblock In {\em Proceedings of the IEEE conference on computer vision and
  pattern recognition}, pages 770--778, 2016.

\bibitem{ibarrondo2018fhe}
Alberto Ibarrondo and Melek {\"O}nen.
\newblock Fhe-compatible batch normalization for privacy preserving deep
  learning.
\newblock In {\em Data Privacy Management, Cryptocurrencies and Blockchain
  Technology}, pages 389--404. Springer, 2018.

\bibitem{ioffe2015batch}
Sergey Ioffe and Christian Szegedy.
\newblock Batch normalization: Accelerating deep network training by reducing
  internal covariate shift.
\newblock In {\em International conference on machine learning}, pages
  448--456. PMLR, 2015.

\bibitem{juvekar2018gazelle}
Chiraag Juvekar, Vinod Vaikuntanathan, and Anantha Chandrakasan.
\newblock $\{$GAZELLE$\}$: A low latency framework for secure neural network
  inference.
\newblock In {\em 27th $\{$USENIX$\}$ Security Symposium ($\{$USENIX$\}$
  Security 18)}, pages 1651--1669, 2018.

\bibitem{kolesnikov2008improved}
Vladimir Kolesnikov and Thomas Schneider.
\newblock Improved garbled circuit: Free xor gates and applications.
\newblock In {\em International Colloquium on Automata, Languages, and
  Programming}, pages 486--498. Springer, 2008.

\bibitem{krishnamoorthi2018quantizing}
Raghuraman Krishnamoorthi.
\newblock Quantizing deep convolutional networks for efficient inference: A
  whitepaper.
\newblock {\em arXiv preprint arXiv:1806.08342}, 2018.

\bibitem{krizhevsky2012imagenet}
Alex Krizhevsky, Ilya Sutskever, and Geoffrey~E Hinton.
\newblock Imagenet classification with deep convolutional neural networks.
\newblock {\em Advances in neural information processing systems},
  25:1097--1105, 2012.

\bibitem{lee2018snip}
Namhoon Lee, Thalaiyasingam Ajanthan, and Philip~HS Torr.
\newblock Snip: Single-shot network pruning based on connection sensitivity.
\newblock {\em arXiv preprint arXiv:1810.02340}, 2018.

\bibitem{liu2017oblivious}
Jian Liu, Mika Juuti, Yao Lu, and Nadarajah Asokan.
\newblock Oblivious neural network predictions via minionn transformations.
\newblock In {\em Proceedings of the 2017 ACM SIGSAC Conference on Computer and
  Communications Security}, pages 619--631, 2017.

\bibitem{mishra2020delphi}
Pratyush Mishra, Ryan Lehmkuhl, Akshayaram Srinivasan, Wenting Zheng, and
  Raluca~Ada Popa.
\newblock Delphi: A cryptographic inference service for neural networks.
\newblock In {\em 29th $\{$USENIX$\}$ Security Symposium ($\{$USENIX$\}$
  Security 20)}, pages 2505--2522, 2020.

\bibitem{mohassel2017secureml}
Payman Mohassel and Yupeng Zhang.
\newblock Secureml: A system for scalable privacy-preserving machine learning.
\newblock In {\em 2017 IEEE Symposium on Security and Privacy (SP)}, pages
  19--38. IEEE, 2017.

\bibitem{nair2010rectified}
Vinod Nair and Geoffrey~E Hinton.
\newblock Rectified linear units improve restricted boltzmann machines.
\newblock In {\em Icml}, 2010.

\bibitem{ophelib}
OPHELib.
\newblock Industrial software systems at abb corporate research, 2010.

\bibitem{orlandi2007oblivious}
Claudio Orlandi, Alessandro Piva, and Mauro Barni.
\newblock Oblivious neural network computing via homomorphic encryption.
\newblock {\em EURASIP Journal on Information Security}, 2007:1--11, 2007.

\bibitem{paillier1999public}
Pascal Paillier.
\newblock Public-key cryptosystems based on composite degree residuosity
  classes.
\newblock In {\em International conference on the theory and applications of
  cryptographic techniques}, pages 223--238. Springer, 1999.

\bibitem{qin2020binary}
Haotong Qin, Ruihao Gong, Xianglong Liu, Xiao Bai, Jingkuan Song, and Nicu
  Sebe.
\newblock Binary neural networks: A survey.
\newblock {\em Pattern Recognition}, 105:107281, 2020.

\bibitem{rastegari2016xnor}
Mohammad Rastegari, Vicente Ordonez, Joseph Redmon, and Ali Farhadi.
\newblock Xnor-net: Imagenet classification using binary convolutional neural
  networks.
\newblock In {\em European conference on computer vision}, pages 525--542.
  Springer, 2016.

\bibitem{riazi2019xonn}
M~Sadegh Riazi, Mohammad Samragh, Hao Chen, Kim Laine, Kristin Lauter, and
  Farinaz Koushanfar.
\newblock $\{$XONN$\}$: Xnor-based oblivious deep neural network inference.
\newblock In {\em 28th $\{$USENIX$\}$ Security Symposium ($\{$USENIX$\}$
  Security 19)}, pages 1501--1518, 2019.

\bibitem{rouhani2018deepsecure}
Bita~Darvish Rouhani, M~Sadegh Riazi, and Farinaz Koushanfar.
\newblock Deepsecure: Scalable provably-secure deep learning.
\newblock In {\em Proceedings of the 55th Annual Design Automation Conference},
  pages 1--6, 2018.

\bibitem{sanyal2018tapas}
Amartya Sanyal, Matt Kusner, Adria Gascon, and Varun Kanade.
\newblock Tapas: Tricks to accelerate (encrypted) prediction as a service.
\newblock In {\em International Conference on Machine Learning}, pages
  4490--4499. PMLR, 2018.

\bibitem{schroff2015facenet}
Florian Schroff, Dmitry Kalenichenko, and James Philbin.
\newblock Facenet: A unified embedding for face recognition and clustering.
\newblock In {\em Proceedings of the IEEE conference on computer vision and
  pattern recognition}, pages 815--823, 2015.

\bibitem{simonyan2014very}
Karen Simonyan and Andrew Zisserman.
\newblock Very deep convolutional networks for large-scale image recognition.
\newblock {\em arXiv preprint arXiv:1409.1556}, 2014.

\bibitem{tramer2019slalom}
Florian Tramèr and Dan Boneh.
\newblock Slalom: Fast, verifiable and private execution of neural networks in
  trusted hardware, 2019.

\bibitem{yang2017designing}
Tien-Ju Yang, Yu-Hsin Chen, and Vivienne Sze.
\newblock Designing energy-efficient convolutional neural networks using
  energy-aware pruning.
\newblock In {\em Proceedings of the IEEE Conference on Computer Vision and
  Pattern Recognition}, pages 5687--5695, 2017.

\bibitem{zhang2018lq}
Dongqing Zhang, Jiaolong Yang, Dongqiangzi Ye, and Gang Hua.
\newblock Lq-nets: Learned quantization for highly accurate and compact deep
  neural networks.
\newblock In {\em Proceedings of the European conference on computer vision
  (ECCV)}, pages 365--382, 2018.

\bibitem{zhang2021gala}
Qiao Zhang, Chunsheng Xin, and Hongyi Wu.
\newblock Gala: Greedy computation for linear algebra in privacy-preserved
  neural networks.
\newblock {\em arXiv preprint arXiv:2105.01827}, 2021.

\end{thebibliography}

%%%%%%%%%%%%%%%%%%%%%%%%%%%%%%%%%%%%%%%%%%%%%%%%%%%%%%%%%%%%%%%%%%%%%%%%%%%%%%%%
\end{document}